\documentclass[11pt,a4paper]{article}

\usepackage[utf8]{inputenc}
\usepackage[english]{babel}
\usepackage{amsmath, amssymb, amsthm}
\usepackage{graphicx}
\usepackage{hyperref}
\usepackage{geometry}
\usepackage{cite}
\usepackage{booktabs}
\usepackage{caption}
\usepackage{subcaption} 
\usepackage{orcidlink}
\usepackage{algorithm}
\usepackage{algorithmicx}
\usepackage{algpseudocode} 

\usepackage{algorithm}
\usepackage{algpseudocode}

\title{\textbf{Raising the Bar: An Asymptotic Comparison of Classical and Quantum Shortest Path Algorithms}}

\author{
  Phuc Hao Do$^{1}$ \\
  \texttt{do.hf@sut.ru,\orcidlink{0000-0003-0645-0021}} \\
  \and
  Tran Duc Le$^{2}$ \\
  \texttt{let@uwstout.edu,\orcidlink{0000-0003-3735-0314}} \\
}

\date{
  $^{1}$Department of Telecommunication Engineering, Bonch-Bruevich St. Petersburg State University of Telecommunications \\
  $^{2}$Mathematics, Statistics and Computer Science, University of Wisconsin–Stout \\[2ex]
}

\begin{document}

\maketitle

\begin{abstract}
The Single-Source Shortest Path (SSSP) problem is a cornerstone of computer science with vast applications, for which Dijkstra's algorithm has long been the classical baseline. While various quantum algorithms have been proposed, their performance has typically been benchmarked against this decades-old approach. This landscape was recently reshaped by the introduction of a new classical algorithm by Duan et al. with a complexity of $O(m \cdot (\log n)^{2/3})$. This development necessitates a re-evaluation of the quantum advantage narrative for SSSP. In this paper, we conduct a systematic theoretical comparison of modern quantum and classical SSSP algorithms in light of this new classical frontier. Through an analysis of their theoretical cost functions, we illustrate how their relative scaling compares across scenarios that vary in graph density and path length. Our analysis suggests a nuanced picture: sophisticated quantum algorithms, such as the one by Wesolowski and Piddock, can exhibit more favorable asymptotic scaling, but only in regimes characterized by short solution paths. Conversely, for problems involving long paths, state-of-the-art classical algorithms appear to maintain a scaling advantage. Our work provides an updated perspective for future quantum algorithm development and underscores that the pursuit of quantum advantage is a dynamic race where the classical goalposts are continually shifting.
\end{abstract}


\section{Introduction}
\label{sec:introduction}

The Single-Source Shortest Path (SSSP) \cite{forster2018faster} problem stands as a cornerstone of modern computer science and algorithmic theory. Its applications are ubiquitous, forming the computational backbone of critical systems that range from the internet's core routing protocols, such as OSPF, to complex logistical optimizations and large-scale analyses in computational biology. For more than six decades, the primary benchmark for solving SSSP on graphs with non-negative edge weights has been Dijkstra's celebrated algorithm~\cite{dijkstra1959note}. The algorithm's well-established time complexity, optimized to $O(m + n \log n)$ using a Fibonacci heap, has not only provided a practical solution for countless problems but has also served as a fixed point of comparison against which new algorithmic innovations are measured.

With the advent of quantum computing, this classical benchmark became a target for disruption. The promise of quantum speedups spurred the development of a new class of algorithms designed to solve SSSP more efficiently \cite{aaa}. Early forays into this domain often adopted a hybrid strategy, leveraging foundational quantum subroutines like Grover's search algorithm~\cite{grover1996fast} to accelerate classically intensive steps, such as finding a minimum-distance vertex. More recently, the field has matured, yielding sophisticated, quantum-native approaches based on quantum walks and intricate divide-and-conquer strategies, exemplified by the work of Wesolowski and Piddock~\cite{wesolowski2024advances}. Despite their diversity, these quantum endeavors have been united by a common thread: their claims of advantage have historically been framed in juxtaposition to the venerable, yet aging, Dijkstra-era baseline.

This long-standing paradigm, however, was recently altered by a significant achievement in classical algorithmics. Duan et al.~\cite{duan2025breaking} introduced a deterministic classical algorithm for SSSP that successfully breaks the theoretical sorting barrier, long considered an inherent bottleneck. Their work establishes a new state-of-the-art with a time complexity of $O(m \cdot (\log n)^{2/3})$. This development is a notable development, redrawing the classical performance frontier. Consequently, it redefines the measure of success for any competing quantum algorithm \cite{chuang1998experimental} and raises an immediate, critical question: how do existing quantum algorithms, once considered promising, compare against this new and more formidable classical competitor?

To date, this question has remained unanswered. The prevailing narrative of quantum advantage for SSSP is still largely anchored in an outdated context, a race against a benchmark that has now been surpassed. This leaves a gap in our understanding of the true potential of quantum computing \cite{rietsche2022quantum} for this fundamental problem. This paper aims to help bridge that gap. We undertake a systematic, theoretical analysis to confront modern quantum algorithms with this new classical frontier, thereby providing a contemporary assessment of the algorithms' comparative asymptotic scaling.

To achieve this, our work unfolds through a structured investigation. We begin by surveying the leading classical and quantum algorithms \cite{mosca2009quantum}, establishing a clear and current set of baselines for comparison. Building on this foundation, we construct a framework for an asymptotic cost analysis designed to compare the theoretical complexities of these algorithms across a spectrum of graph structures. This framework allows us to explore the impact of two critical parameters: the graph's edge density, which distinguishes sparse real-world networks from highly interconnected \cite{ghavasieh2024diversity} ones, and the geometric nature of the solution path itself, particularly its length. By analyzing the interplay of these factors, we illustrate the problem regimes - which we term "Quantum Advantage Zones" \cite{fedorov2022quantum} for the purpose of this discussion - where the asymptotic scaling of quantum approaches suggests a significant performance advantage. Finally, we synthesize these findings to discuss their implications, arguing that the pursuit of quantum advantage \cite{do2025challenges} is not a static goal but a dynamic race against a continuously advancing classical frontier, a realization that carries consequences for the future of quantum algorithm design for graph problems and beyond.

The remainder of this paper is structured to guide the reader through this investigation. Section~\ref{sec:preliminaries} provides the necessary background on the algorithms under study. Section~\ref{sec:methodology} details our analytical methodology. In Section~\ref{sec:results}, we present and analyze the results of our comparative analysis. The broader implications of these findings are discussed in Section~\ref{sec:discussion}, before we conclude the paper in Section~\ref{sec:conclusion}.

\section{Preliminaries and Baselines}
\label{sec:preliminaries}

To establish a clear foundation for our comparative analysis, this section delves into the algorithmic underpinnings of the key contenders in the shortest path problem. We begin by formally defining the problem and reviewing the classical algorithms that have historically defined and recently redefined the performance benchmarks. Subsequently, we survey the primary paradigms of quantum approaches, detailing the distinct strategies they employ to seek a computational advantage.

\subsection{Classical Shortest Path Algorithms}

The SSSP problem is formally defined on a weighted \cite{bapat2012weighted} directed graph $G = (V, E)$, where $V$ is the set of $n$ vertices and $E$ is the set of $m$ edges. Each edge $(u, v) \in E$ is associated with a non-negative weight $w(u, v)$. Given a designated source vertex $s \in V$, the objective is to compute the length of the shortest path from $s$ to every other vertex $v \in V$.

\subsubsection{The Dijkstra-era Baseline}
For over half a century, the canonical solution to this problem has been Dijkstra's algorithm~\cite{dijkstra1959note}. It operates on a greedy principle, iteratively exploring the graph by selecting the unvisited vertex with the smallest known distance from the source. The efficiency of this selection process is paramount and hinges on the implementation of a priority queue data structure. While a simple binary heap yields a respectable complexity, the theoretical standard is achieved using a Fibonacci heap~\cite{fredman1987fibonacci}, which optimizes the key operations to yield an overall time complexity of $O(m + n \log n)$. This complexity has been remarkably resilient, serving for decades as the effective \textit{classical baseline} - a formidable benchmark that any aspiring algorithm, whether classical or quantum, must surpass to claim superiority.

\subsubsection{The New Classical Frontier}
The prevailing assumption that the $O(n \log n)$ term represented an unavoidable "sorting barrier" was recently overturned by a notable result from Duan et al.~\cite{duan2025breaking}. Their deterministic algorithm introduces a novel approach that circumvents the need for a traditional priority queue structure, thereby breaking this long-standing theoretical bottleneck. The result is a significantly improved time complexity of $O(m \cdot (\log n)^{2/3})$. This achievement establishes a \textit{new classical frontier}, raising the bar for what constitutes a high-performance SSSP algorithm. It is against this new, more aggressive benchmark that the asymptotic scaling of modern quantum algorithms should be measured.

\subsection{A Survey of Quantum Approaches}

The quest for a quantum solution to SSSP \cite{van2018improvements} has evolved along two distinct conceptual paths. The first involves leveraging general-purpose quantum subroutines to accelerate parts of a classical algorithmic structure. The second, more ambitious path aims to design novel, quantum-native algorithms that solve the problem in a fundamentally different way.

\subsubsection{Grover-based Search Acceleration: A Hybrid Approach}
One of the earliest and most intuitive strategies for a quantum SSSP speedup is a hybrid quantum-classical approach centered on Grover's celebrated search algorithm~\cite{grover1996fast}. The core concept is to retain the iterative structure of an algorithm like Dijkstra's but to offload the most computationally intensive step - finding the minimum-distance vertex among all unvisited candidates - to a quantum processor. This task is an instance of the unstructured minimum-finding problem, for which quantum computers can offer a quadratic speedup over classical brute-force search~\cite{durr1996quantum}, requiring only $O(\sqrt{k})$ queries to find the minimum of $k$ items.

When integrated into an overall shortest path algorithm, this quantum enhancement typically leads to a time complexity of $O(\sqrt{n} \cdot m)$~\cite{dorn2007quantum}. While this complexity represents a clear polynomial speedup over the most naive classical implementations (e.g., $O(n \cdot m)$), its competitiveness against highly optimized classical algorithms like Dijkstra's with a Fibonacci heap is not guaranteed, especially on sparse graphs where the dependence on $m$ is punishing. Nevertheless, its conceptual simplicity and its status as a foundational approach make it a crucial reference point for understanding the evolution of quantum graph algorithms.

\subsubsection{Divide and Conquer via Quantum Walks: A Quantum-Native Strategy}
A more sophisticated and structurally distinct paradigm is presented by Wesolowski and Piddock~\cite{wesolowski2024advances}. Their approach moves beyond simple subroutine acceleration and constitutes a quantum-native strategy. The algorithm is built upon a quantum divide-and-conquer framework that recursively breaks the problem into smaller pieces. Instead of incrementally building paths from a source vertex in a greedy fashion, this method leverages quantum phenomena to explore the graph's global structure and construct the solution.

The resulting time complexity is $\tilde{O}(l \cdot \sqrt{m})$. The $\tilde{O}$ (soft-O) notation is used to omit polylogarithmic factors in $n$ and $m$. While these factors are critically important for determining precise performance crossover points, our high-level analysis will focus on the dominant scaling terms to compare the fundamental structure of the complexities. The most critical feature is the complexity's linear dependence on $l$, the total weight (or length) of the final shortest path \cite{orda1990shortest}, and its square-root dependence on the number of edges, $m$. This creates a distinct performance profile where efficiency is not primarily dictated by the size of the graph ($n$ or $m$), but rather by the geometric properties of the solution path itself. This suggests that its asymptotic scaling will be favorable in scenarios where the shortest path is structurally "short," but may be unfavorable when the solution requires a long route. This nuanced, input-dependent behavior makes it a compelling candidate in the modern quest for a demonstrable \cite{lorenz2024quest} quantum advantage.

\section{Methodology for Asymptotic Analysis}
\label{sec:methodology}

To theoretically compare the performance scaling of classical \cite{pathak2024new} and quantum SSSP algorithms, our study employs an analysis of their published theoretical time complexities. This approach is chosen to focus purely on the fundamental asymptotic behavior of the algorithms. By abstracting away implementation-specific overheads, constant factors, and hardware dependencies, we can compare the mathematical structure of the cost functions as the problem size scales. This analysis rests on two components: the formulation of the algorithmic cost functions and a set of illustrative scenarios designed to test these functions under diverse structural assumptions.

\subsection{Algorithmic Cost Functions}
\label{subsec:cost_functions}

The foundation of our analysis is a set of cost functions, each denoted by $C(\cdot)$, which represent the established theoretical time complexity for the four algorithms under investigation. These functions take graph parameters - the number of vertices ($n$), edges ($m$), and where applicable, path length ($l$) - as input. It is critical to note that they output a unitless value representing the theoretical computational cost. Our analysis compares these values directly, which is equivalent to assuming all hidden constant factors are 1. This simplification allows for a direct comparison of the algorithms' scaling behavior but does not constitute a prediction of concrete, real-world performance. These models are derived from the literature discussed in Section~\ref{sec:preliminaries}.

\begin{itemize}

\item \textbf{Dijkstra's Algorithm (Classical Baseline):} The cost is modeled after the implementation using a Fibonacci heap.
\begin{equation}
    C_{\text{Dijkstra}}(n, m) = m + n \log_2 n
    \label{eq:dijkstra}
\end{equation}

\item \textbf{Duan et al.'s Algorithm (New Classical Frontier):} This function encapsulates the sub-logarithmic improvement defining the new classical state-of-the-art.
\begin{equation}
    C_{\text{Duan}}(n, m) = m \cdot (\log_2 n)^{2/3}
    \label{eq:duan}
\end{equation}

\item \textbf{Grover-based Quantum Algorithm:} This cost function reflects the quadratic speedup from quantum search applied within a classical iterative structure.
\begin{equation}
    C_{\text{Grover}}(n, m) = \sqrt{n} \cdot m
    \label{eq:grover}
\end{equation}

\item \textbf{Wesolowski et al.'s Quantum Algorithm:} This model depends on the solution geometry ($l$). Consistent with our discussion in Section~\ref{sec:preliminaries}, this model omits the polylogarithmic factors hidden by the $\tilde{O}$ notation to focus the analysis on the dominant scaling terms.
\begin{equation}
    C_{\text{Wesolowski}}(n, m, l) = l \cdot \sqrt{m}
    \label{eq:wesolowski}
\end{equation}

\end{itemize}

\subsection{Analytical Scenarios}
\label{subsec:scenarios}

Our analytical framework is designed to systematically compare the algorithms' scaling under conditions that represent archetypal graph structures. We analyze a $2 \times 2$ matrix of scenarios defined by two axes: \textbf{graph density} and \textbf{path length}. These scenarios are not exhaustive but were chosen to illustrate performance trade-offs under common and contrasting conditions. The procedure is detailed in Algorithm~\ref{alg:simulation}.

\begin{algorithm}[h]
\caption{Asymptotic Cost Comparison Framework}
\footnotesize
\label{alg:simulation}
\begin{algorithmic}[1]
\State \textbf{Input:} A set of vertex counts $N_{range}$ to analyze.
\State \textbf{Input:} A set of scenarios $S$, where each $s \in S$ is a tuple $(name, m(n), l(n))$.
\State \textbf{Output:} A set of figures and tables comparing theoretical algorithm costs.

\ForAll{scenario $s \in S$}
    \State Initialize data structures $Results_s$
    \ForAll{$n \in N_{range}$}
        \State $m \gets \text{evaluate } m(n) \text{ from } s$
        \State $l \gets \text{evaluate } l(n) \text{ from } s$
        
        \State $cost_D \gets C_{\text{Dijkstra}}(n, m)$ \Comment{Eq.~\ref{eq:dijkstra}}
        \State $cost_{Du} \gets C_{\text{Duan}}(n, m)$ \Comment{Eq.~\ref{eq:duan}}
        \State $cost_G \gets C_{\text{Grover}}(n, m)$ \Comment{Eq.~\ref{eq:grover}}
        \State $cost_W \gets C_{\text{Wesolowski}}(n, m, l)$ \Comment{Eq.~\ref{eq:wesolowski}}
        
        \State Store $(cost_D, cost_{Du}, cost_G, cost_W)$ in $Results_s$ for the current $n$.
    \EndFor
    \State Generate plot and table for scenario $s$ using $Results_s$.
\EndFor
\end{algorithmic}
\end{algorithm}

\textbf{Graph Density Axis:} This axis investigates how scaling is affected by graph connectivity. We define two representative regimes:
\begin{itemize}
    \item \textbf{Sparse Graphs:} We set $m = 10n$. This linear relationship is a common model for geographically or physically constrained networks, such as road systems or the internet backbone, where the number of connections per node is relatively small.
    \item \textbf{Dense Graphs:} We set $m = n^2 / 100$. This quadratic scaling is chosen to model highly interconnected networks where a significant fraction of all possible edges exist, such as in dense social communities or protein-interaction networks.
\end{itemize}

\textbf{Path Length Axis:} This axis probes the sensitivity of the Wesolowski et al. algorithm to its path length parameter, $l$.
\begin{itemize}
    \item \textbf{Short Paths:} We define this scenario with $l = (\log_2 n)^2$. A polylogarithmic path length is characteristic of "small-world" networks, where paths between nodes are typically short relative to the network size.
    \item \textbf{Long Paths:} We define this with $l = n/10$. This linear relationship models a challenging case where the shortest path is long, requiring traversal of a significant fraction of the graph's vertices.
\end{itemize}

By applying our analysis (Algorithm~\ref{alg:simulation}) to all four combinations of these parameters, we generate comparative plots that illustrate the different scaling behaviors. These plots help to identify the conditions under which each algorithm's complexity formula yields a more favorable scaling, suggesting the boundaries of regimes where quantum approaches may hold an asymptotic advantage. All calculations and plotting were implemented in Python using the NumPy and Matplotlib libraries.

\section{Results and Analysis}
\label{sec:results}

This section presents the results of our comparative analysis. We systematically examine the outputs of the cost functions, guided by the methodology established in Section~\ref{sec:methodology}. Our analysis first explores the scaling behavior on sparse graphs - an important analogue for many real-world networks - before turning to the distinct conditions posed by dense graphs. We then synthesize these observations to illustrate the conditions that influence the potential for a quantum advantage in SSSP.

\subsection{Scaling on Sparse Graphs: The Decisive Role of Path Length}

In the domain of sparse graphs, where the number of edges scales linearly with vertices ($m=10n$), our analysis reveals a clear difference in the scaling behavior of the algorithms. The determining factor is not the size of the graph itself, but the geometric character of the solution path as defined in the scenario.

\begin{figure}[h]
    \centering
    \includegraphics[width=\columnwidth]{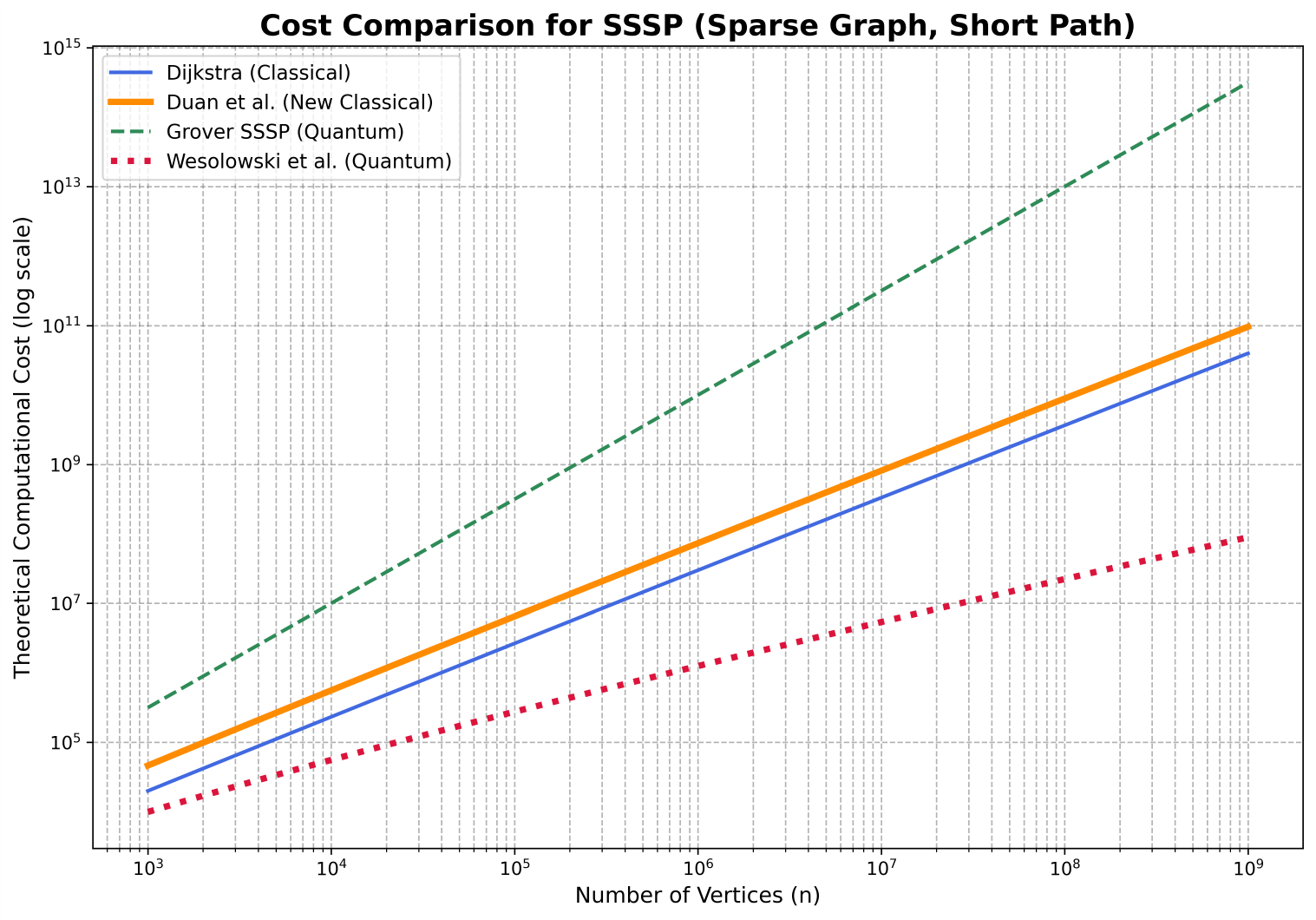}
    \caption{Cost comparison on sparse graphs ($m=10n$) with short path lengths ($l=(\log_2 n)^2$). The Wesolowski et al. algorithm (red, dotted) demonstrates a clear and growing scaling advantage over all classical and quantum counterparts.}
    \label{fig:sparse_short}
\end{figure}

The first scenario, which models short paths where $l=(\log_2 n)^2$, illustrates a compelling case for an asymptotic quantum advantage. As shown in Figure~\ref{fig:sparse_short}, the cost trajectory of the Wesolowski et al. algorithm scales more favorably than its competitors. This visual trend is further detailed by the sample calculations in Table~\ref{tab:sparse_results}. For a graph with $n=10^8$ vertices, the theoretical cost function for the quantum algorithm produces a value more than 150 times lower than that of the best classical alternative. This ratio, derived under the simplifying assumptions outlined in our methodology, grows with $n$, suggesting a fundamental scaling superiority in this regime. In contrast, the Grover-based approach is clearly non-competitive, with a cost value several orders of magnitude higher than the Dijkstra baseline, confirming that its $O(\sqrt{n} \cdot m)$ complexity is ill-suited for sparse graphs.

\begin{table}[h]
    \centering
    \caption{Theoretical Cost Comparison on Sparse Graphs ($m = 10n$).}
    \label{tab:sparse_results}
    \resizebox{\textwidth}{!}{%
    \begin{tabular}{l l rrrr c}
        \toprule
        \textbf{Scenario} & \textbf{Vertices ($n$)} & \textbf{Dijkstra} & \textbf{Duan et al.} & \textbf{Grover SSSP} & \textbf{Wesolowski et al.} & \textbf{Winner} \\
        \midrule
        \textbf{Short Path} & $10^4$ & $2.16 \times 10^5$ & $5.21 \times 10^5$ & $9.01 \times 10^6$ & $\mathbf{5.31 \times 10^4}$ & Wesolowski \\
        \textit{($l = (\log_2 n)^2$)} & $10^6$ & $2.78 \times 10^7$ & $6.83 \times 10^7$ & $9.01 \times 10^9$ & $\mathbf{1.20 \times 10^6}$ & Wesolowski \\
        & $10^8$ & $3.40 \times 10^9$ & $8.28 \times 10^9$ & $9.01 \times 10^{12}$ & $\mathbf{2.14 \times 10^7}$ & Wesolowski \\
        \midrule
        \textbf{Long Path} & $10^4$ & $\mathbf{2.16 \times 10^5}$ & $5.21 \times 10^5$ & $9.01 \times 10^6$ & $2.85 \times 10^5$ & Dijkstra \\
        \textit{($l = n/10$)} & $10^6$ & $\mathbf{2.78 \times 10^7}$ & $6.83 \times 10^7$ & $9.01 \times 10^9$ & $2.85 \times 10^8$ & Dijkstra \\
        & $10^8$ & $\mathbf{3.40 \times 10^9}$ & $8.28 \times 10^9$ & $9.01 \times 10^{12}$ & $2.85 \times 10^{11}$ & Dijkstra \\
        \bottomrule
    \end{tabular}%
    }
    \footnotesize{\\ Costs are the unitless outputs of the theoretical complexity functions, assuming all hidden constants are 1. Winner is determined by the lowest cost. Bold values indicate the winning cost.}
\end{table}

\begin{figure}[t]
    \centering
    \includegraphics[width=\columnwidth]{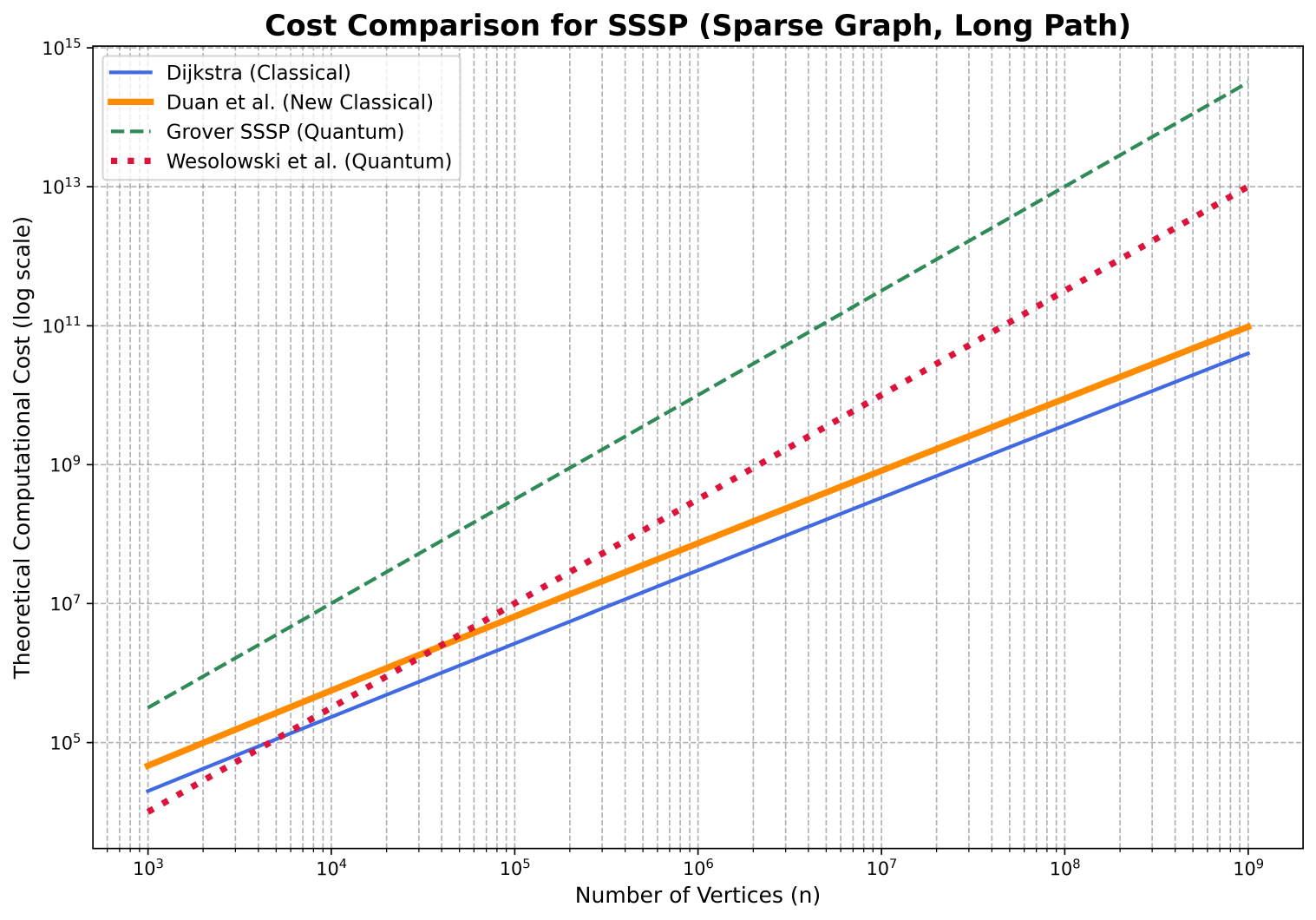}
    \caption{Cost comparison on sparse graphs ($m=10n$) with long path lengths ($l=n/10$). The classical algorithms (blue and orange) show more favorable scaling, as the quantum algorithms' costs are higher due to dependencies on $l$ and $n$.}
    \label{fig:sparse_long}
\end{figure}

However, this favorable scaling for the quantum algorithm is reversed in the long-path scenario, where $l=n/10$. Figure~\ref{fig:sparse_long} depicts this reversal in relative scaling. As is expected from its cost function, the linear dependence on path length becomes a significant performance liability for the Wesolowski et al. algorithm. As $l$ grows linearly with $n$, its complexity, $l \cdot \sqrt{m} \approx (n/10) \cdot \sqrt{10n} = O(n^{1.5})$, scales unfavorably. The bottom section of Table~\ref{tab:sparse_results} illustrates this trend: the classical algorithms, with their $O(n \log n)$ or similar scaling, become orders of magnitude more efficient. This result highlights a critical boundary condition for the applicability of this quantum approach, illustrating that its asymptotic advantage is highly dependent on the problem structure.

\subsection{Scaling on Dense Graphs: The Interplay of Density and Geometry}

Turning to dense graphs, where $m=n^2/100$, the large number of edges becomes a primary bottleneck for algorithms with a strong dependence on $m$. Our analysis in this regime reveals a more complex interplay between algorithmic structure and the problem parameters.

\begin{figure}[h]
    \centering
    \includegraphics[width=\columnwidth]{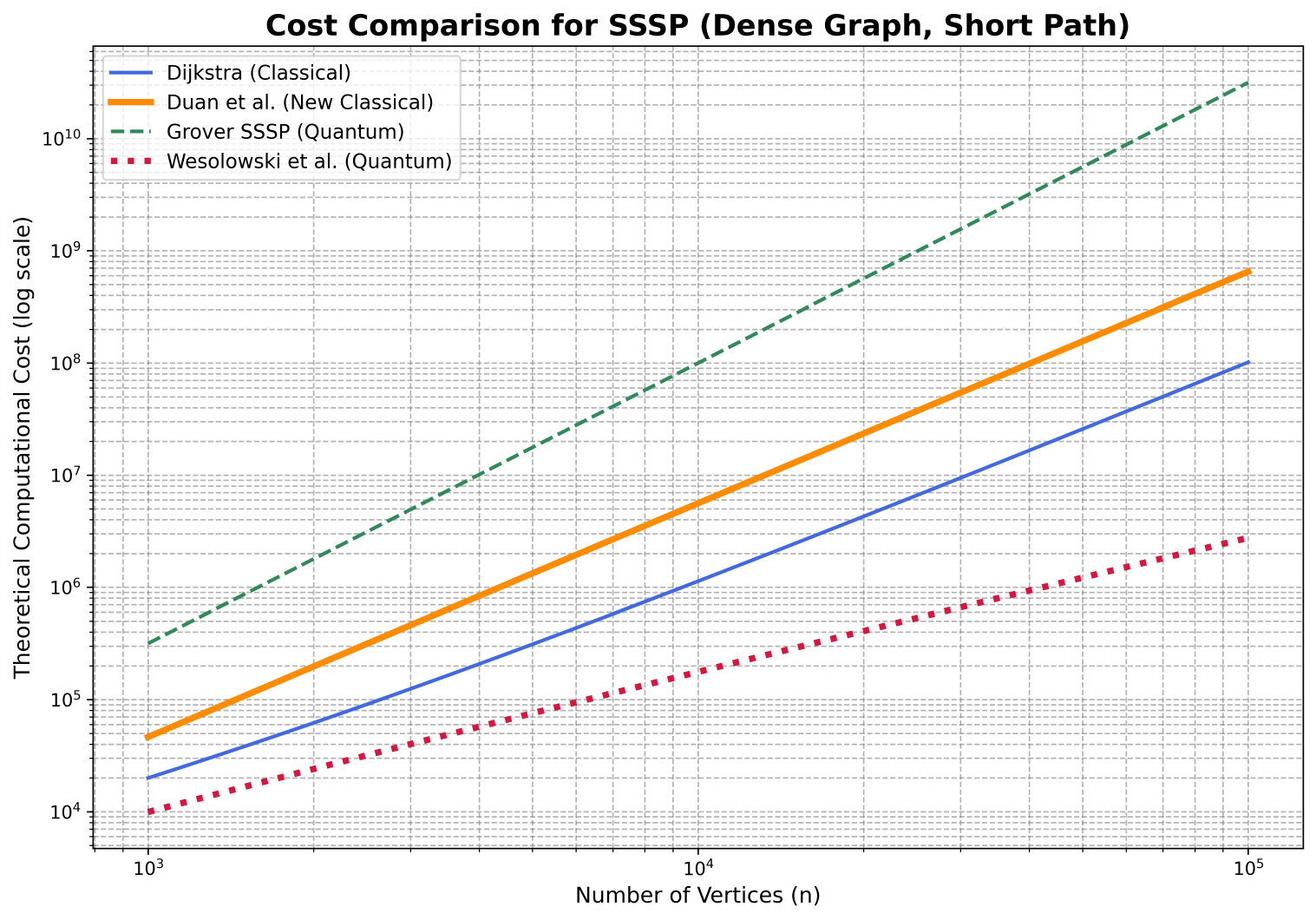}
    \caption{Cost comparison on dense graphs ($m=n^2/100$) with short path lengths. The asymptotic advantage of the Wesolowski et al. algorithm persists due to its sublinear scaling with $m$.}
    \label{fig:dense_short}
\end{figure}

In the short-path scenario, the asymptotic advantage for the Wesolowski et al. algorithm persists and becomes more pronounced. As shown in Figure~\ref{fig:dense_short}, its reliance on $\sqrt{m}$ provides a crucial edge over the classical methods, whose costs are driven by the now-quadratic term $m$. This highlights the algorithm's structural resilience to high connectivity.

\begin{figure}[h]
    \centering
    \includegraphics[width=\columnwidth]{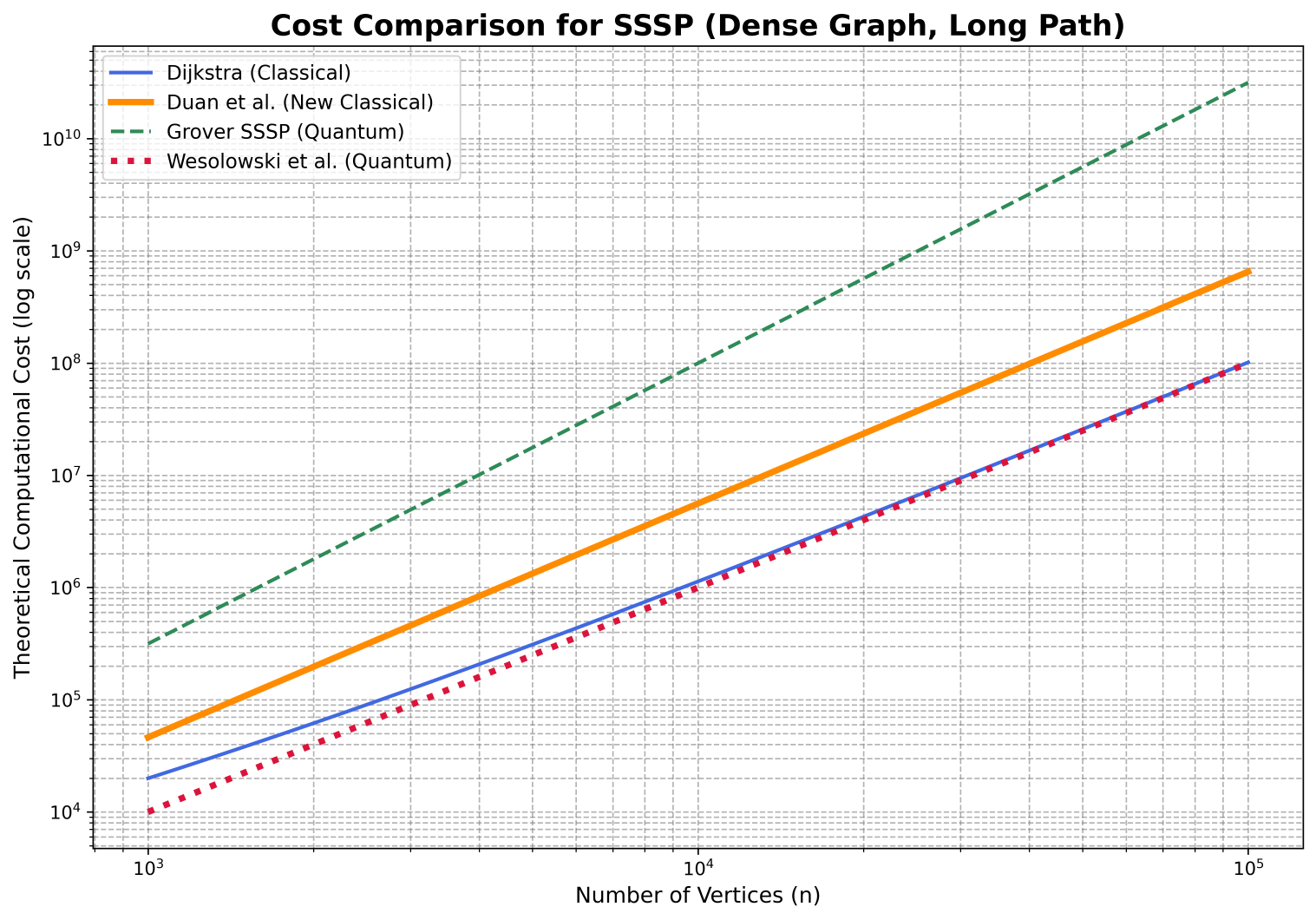}
    \caption{Cost comparison on dense graphs ($m=n^2/100$) with long path lengths. The scaling behaviors are very close, indicating a potential crossover point where the asymptotic advantage becomes marginal.}
    \label{fig:dense_long}
\end{figure}

The most balanced comparison in our study emerges from the dense-graph, long-path scenario, depicted in Figure~\ref{fig:dense_long}. Here, a trade-off between two factors becomes apparent: the quantum algorithm's more favorable scaling with high edge density versus its unfavorable scaling with long path lengths. The result is a close comparison of the cost function values. The calculation for $n=10^4$ in Table~\ref{tab:speedup_analysis} shows the values to be nearly identical, with a cost ratio of just $\approx 1.13$. This region represents a potential "crossover point" in the asymptotic performance landscape, where the optimal algorithmic choice is highly sensitive to the precise parameters of the problem instance.

\subsection{Illustrative Ratios of Asymptotic Costs}

To synthesize the preceding observations, we examine the ratio of the best-performing classical cost function to the Wesolowski et al. cost function in each of our four scenarios. The results, consolidated in Table~\ref{tab:speedup_analysis}, provide a set of illustrative comparisons for the "Quantum Advantage Zone." It is important to reiterate that these ratios are not predictions of real-world speedup but are comparisons of the unitless outputs of the simplified cost models.

\begin{table}[h]
    \centering
    \caption{Illustrative Ratios of Theoretical Costs (Wesolowski et al. vs. Best Classical).}
    \label{tab:speedup_analysis}
    \resizebox{\textwidth}{!}{%
    \begin{tabular}{l l l c c c c}
        \toprule
        \textbf{Graph Type} & \textbf{Path Length} & \textbf{Vertices ($n$)} & \textbf{Best Classical Cost (A)} & \textbf{Wesolowski Cost (B)} & \textbf{Cost Ratio (A/B)} & \textbf{Asymptotic Advantage?} \\
        \midrule
        \textbf{Sparse} & \textbf{Short} & $10^4$ & $2.16 \times 10^5$ & $5.31 \times 10^4$ & $\approx 4.1$ & Quantum \\
        ($m=10n$) & ($l=(\log_2 n)^2$) & $10^6$ & $2.78 \times 10^7$ & $1.20 \times 10^6$ & $\approx 23.2$ & Quantum \\
        & & $10^8$ & $3.40 \times 10^9$ & $2.14 \times 10^7$ & $\approx 158.9$ & Quantum \\
        \cmidrule{2-7}
        & \textbf{Long} & $10^4$ & $2.16 \times 10^5$ & $2.85 \times 10^5$ & $\approx 0.76$ & Classical \\
        & ($l=n/10$) & $10^6$ & $2.78 \times 10^7$ & $2.85 \times 10^8$ & $\approx 0.10$ & Classical \\
        \midrule
        \textbf{Dense} & \textbf{Short} & $10^4$ & $1.08 \times 10^6$ & $1.72 \times 10^5$ & $\approx 6.3$ & Quantum \\
        ($m=n^2/100$) & ($l=(\log_2 n)^2$) & & & & & \\
        \cmidrule{2-7}
        & \textbf{Long} & $10^4$ & $1.08 \times 10^6$ & $9.55 \times 10^5$ & $\approx 1.13$ & Marginal \\
        & ($l=n/10$) & & & & & \\
        \bottomrule
    \end{tabular}%
    }
    \footnotesize{\\ Cost Ratio is (Best Classical Cost) / (Wesolowski et al. Cost). A ratio $>$ 1 suggests a quantum asymptotic advantage. The best classical algorithm was Dijkstra in all cases shown. These values are illustrative of scaling behavior, not predictive of real-world speedups.}
\end{table}

The table suggests the conditions under which a quantum algorithm might have an asymptotic advantage. A significant and growing cost ratio is observed whenever the path length ($l$) is logarithmically small relative to the graph size. This trend is most pronounced in the sparse-graph, short-path case, where the cost function ratio reaches nearly 160 for large $n$, indicating a strong asymptotic advantage under these conditions. Conversely, when the path length is linear in $n$, the quantum algorithm consistently exhibits less favorable scaling, leading to what could be called a "Classical Advantage Zone." The marginal result in the dense, long-path scenario is particularly illuminating, as it highlights the delicate balance of parameters that defines the boundary where the asymptotic advantage transitions between classical and quantum approaches.

\section{Discussion and Implications}
\label{sec:discussion}

The results of our analysis illustrate the trade-offs in the competitive landscape for shortest path algorithms. Our theoretical comparison suggests that the notion of "quantum advantage" for this problem is not a binary proposition but a conditional state, highly dependent on the structural and geometric properties of the problem instance. In this section, we discuss the key takeaways from our analysis and their implications for future quantum algorithm design and high-performance computing.

\subsection{The Grover Barrier: A Heuristic for Simple Speedups}

A clear conclusion from our analysis is that generic, Grover-based search acceleration shows significant limitations for achieving a meaningful asymptotic advantage in SSSP. Across every scenario, the $O(\sqrt{n} \cdot m)$ complexity of this hybrid approach resulted in less favorable scaling not only compared to the new classical frontier established by Duan et al., but even to the decades-old Dijkstra baseline. This outcome illustrates what can be termed the "Grover Barrier" for this class of problems: where highly optimized classical algorithms have complexities with polylogarithmic dependencies (e.g., factors of $\log n$), a quadratic speedup applied to a polynomial factor (i.e., turning $n$ into $\sqrt{n}$) is often mathematically insufficient to offer a competitive edge. The classical complexities are dominated by terms that Grover's search does not address. This suggests that future progress in quantum graph algorithms will likely require moving beyond black-box applications of Grover's algorithm to classical templates, demanding designs that exploit quantum phenomena in a more problem-specific manner.

\subsection{The Path Length Barrier: A Dependency on Solution Geometry}

While the Grover-based approach shows poor scaling, our analysis highlights the algorithm by Wesolowski and Piddock~\cite{wesolowski2024advances} as a genuinely competitive quantum contender in certain regimes. However, its strength introduces a critical dependency that is central to our analysis: the path length, $l$. We use the term "Path Length Barrier" as a conceptual label for this dependency, which separates problems where quantum approaches show favorable scaling from those where classical methods appear to remain superior. Our analysis illustrates two such regimes. The first is a potential Quantum Advantage Zone, characterized by logarithmically short paths ($l = O(\text{polylog}(n))$). Here, the quantum algorithm's scaling is highly favorable; as shown in our illustrative calculations, the ratio between the classical and quantum cost functions can exceed 100 for large graphs. This highlights its potential for "small-world" network problems, such as finding degrees of separation in social networks.

Conversely, the second regime is a Classical Advantage Zone, which emerges when the path length is linear in the number of vertices ($l = O(n)$). In this scenario, the asymptotic advantage is reversed, with classical algorithms demonstrating more favorable scaling. This barrier highlights a key challenge for quantum algorithm designers. The objective is not only to mitigate dependencies on the graph's size ($n$ or $m$), but also to design algorithms that are robust to the geometric properties of the solution itself. Taming the dependence on $l$ remains a significant challenge in the field.

\subsection{Crucial Caveats and Methodological Limitations}

Our conclusions must be interpreted through the lens of several significant methodological limitations. Our analysis is purely theoretical, based on a comparison of asymptotic complexities. As stated in our methodology, this approach does not account for several factors critical to any practical implementation.
\begin{itemize}
    \item The Big-O and Tilde-O notations systematically hide constant factors. Our comparison of unitless cost functions is equivalent to assuming these factors are all 1, but in practice, they can be large enough to alter performance outcomes, especially on moderately sized problems.
    \item We do not model the substantial resource overhead associated with fault-tolerant quantum computation. The costs of quantum error correction, state preparation, and gate operations are expected to add significant polynomial overheads not captured in our high-level analysis.
    \item Many advanced quantum graph algorithms, including the one analyzed here, assume the existence of a Quantum Random Access Memory (QRAM), a technology that is itself a major, unsolved engineering challenge.
\end{itemize}
Therefore, this analysis should be viewed as a high-level comparison of scaling behaviors, intended to guide theoretical inquiry rather than to predict the performance of any future physical implementation.

\subsection{Implications for the Future of Quantum Algorithm Design}

The conditional nature of the asymptotic quantum advantage revealed by our study carries strategic implications for future research. Firstly, it underscores the necessity of moving beyond headline complexity figures. A holistic analysis that considers performance across a spectrum of input parameters is essential, as an algorithm that is only superior in a narrow regime may have limited impact. Secondly, our analysis supports the pursuit of "structurally aware" quantum algorithms. The most promising future avenues will likely involve algorithms that leverage quantum mechanics to exploit the global structure of a graph in ways that are inaccessible to classical, iterative methods. The Wesolowski et al. algorithm, which is sensitive to the solution's geometry, is a testament to this principle. Finally, the challenge highlighted by the "Path Length Barrier" - designing algorithms that are robust to the solution's geometry while retaining benefits like sublinear scaling with $m$ - remains a compelling direction for future research.

\subsection{Implications for Practitioners and High-Performance Computing}

For practitioners facing a shortest path problem, our analysis advocates for a nuanced, data-aware approach to algorithm selection. The era of a single "best" algorithm is over; the optimal choice is contingent on domain-specific knowledge of the problem's characteristics. A conceptual guideline emerges from our theoretical comparison: if the problem context suggests that solutions will be structurally "short" - such as finding connections within a local community in a social network - then quantum algorithms of the Wesolowski type represent a compelling future direction. However, if the problem may involve finding long, winding paths across vast networks - as in global logistics planning or VLSI design - then the stability and favorable scaling of state-of-the-art classical algorithms suggest they are currently the more prudent and performant choice.

\section{Conclusion}
\label{sec:conclusion}

In this paper, we addressed the notable shift in the high-performance computing landscape for the SSSP problem, motivated by a significant advance in classical algorithm design that altered the long-standing performance baseline. Our work provides a systematic theoretical comparison of current classical algorithms against modern quantum approaches in this new competitive context.

Our primary contribution is an illustrative analysis of the conditions under which quantum algorithms might hold an asymptotic advantage. Through a comparative analysis of their theoretical complexities, we have shown that the prospect of a quantum speedup for SSSP is not a certainty but is highly contingent on the structural properties of the problem instance. Our analysis suggests that advanced quantum algorithms can exhibit more favorable asymptotic scaling for problems characterized by short path lengths, regardless of graph density. Conversely, for problems involving long paths, the newest generation of classical algorithms appears to maintain a clear advantage in scaling performance. This highlights the "Path Length Barrier" - the dependency on solution geometry - as a key consideration for future quantum algorithm development.

Ultimately, our analysis underscores a vital lesson for the era of quantum computing: the race for computational supremacy is not a static sprint but a dynamic co-evolution between classical and quantum methodologies. As classical algorithms continue to advance, the bar for demonstrating true quantum advantage will continue to rise. Future progress will depend not on generic speedups, but on the discovery of novel quantum strategies that are deeply attuned to the specific structure of the problems they aim to solve.



\section*{Code Availability}
\addcontentsline{toc}{section}{Code Availability}

The complete source code for the comparative analysis presented in this paper is openly available. The repository contains the Python scripts, utilizing the NumPy and Matplotlib libraries, which implement the theoretical cost functions for all classical and quantum algorithms discussed. These scripts are sufficient to reproduce all figures (Figures 1-4) and tables (Tables 1-2) in this study, ensuring the full reproducibility of our calculations and visualizations. The code is hosted on GitHub at the following repository: 
\url{https://github.com/ailabteam/Quantitative-Re-evaluation-of-Quantum-Speedups}.


\begin{thebibliography}{10}

\bibitem{forster2018faster}
S.~Forster and D.~Nanongkai, ``A faster distributed single-source shortest paths algorithm,'' in {\em 2018 IEEE 59th Annual Symposium on Foundations of Computer Science (FOCS)}, pp.~686--697, IEEE, 2018.

\bibitem{dijkstra1959note}
E.~W. Dijkstra, ``A note on two problems in connexion with graphs,'' {\em Numerische mathematik}, vol.~1, no.~1, pp.~269--271, 1959.

\bibitem{aaa}
{Ashvinkumar, Vikrant}, {Bernstein, Aaron}, {Cao, Nairen}, {Grunau, Christoph}, {Haeupler, Bernhard}, {Jiang, Yonggang}, {Nanongkai, Danupon}, and {Su, Hsin-Hao}, ``Parallel, distributed, and quantum exact single-source shortest paths with negative edge weights,'' 2024.

\bibitem{grover1996fast}
L.~K. Grover, ``A fast quantum mechanical algorithm for database search,'' in {\em Proceedings of the twenty-eighth annual ACM symposium on Theory of computing}, pp.~212--219, 1996.

\bibitem{wesolowski2024advances}
A.~Wesolowski and S.~Piddock, ``Advances in quantum algorithms for the shortest path problem,'' {\em arXiv preprint arXiv:2408.10427}, 2024.

\bibitem{duan2025breaking}
R.~Duan, J.~Mao, X.~Mao, X.~Shu, and L.~Yin, ``Breaking the sorting barrier for directed single-source shortest paths,'' in {\em Proceedings of the 57th Annual ACM Symposium on Theory of Computing}, STOC ’25, p.~36–44, ACM, June 2025.

\bibitem{chuang1998experimental}
I.~L. Chuang, L.~M. Vandersypen, X.~Zhou, D.~W. Leung, and S.~Lloyd, ``Experimental realization of a quantum algorithm,'' {\em Nature}, vol.~393, no.~6681, pp.~143--146, 1998.

\bibitem{rietsche2022quantum}
R.~Rietsche, C.~Dremel, S.~Bosch, L.~Steinacker, M.~Meckel, and J.-M. Leimeister, ``Quantum computing,'' {\em Electronic Markets}, vol.~32, no.~4, pp.~2525--2536, 2022.

\bibitem{mosca2009quantum}
M.~Mosca, ``Quantum algorithms,'' in {\em Encyclopedia of Complexity and Systems Science}, pp.~7088--7118, Springer, 2009.

\bibitem{ghavasieh2024diversity}
A.~Ghavasieh and M.~De~Domenico, ``Diversity of information pathways drives sparsity in real-world networks,'' {\em Nature Physics}, vol.~20, no.~3, pp.~512--519, 2024.

\bibitem{fedorov2022quantum}
A.~K. Fedorov, N.~Gisin, S.~M. Beloussov, and A.~I. Lvovsky, ``Quantum computing at the quantum advantage threshold: a down-to-business review,'' {\em arXiv preprint arXiv:2203.17181}, 2022.

\bibitem{do2025challenges}
P.~H. Do and T.~D. Le, ``Challenges in applying variational quantum algorithms to dynamic satellite network routing,'' {\em arXiv preprint arXiv:2508.04288}, 2025.

\bibitem{bapat2012weighted}
R.~Bapat, D.~Kalita, and S.~Pati, ``On weighted directed graphs,'' {\em Linear Algebra and its Applications}, vol.~436, no.~1, pp.~99--111, 2012.

\bibitem{fredman1987fibonacci}
M.~L. Fredman and R.~E. Tarjan, ``Fibonacci heaps and their uses in improved network optimization algorithms,'' {\em Journal of the ACM (JACM)}, vol.~34, no.~3, pp.~596--615, 1987.

\bibitem{van2018improvements}
J.~Van~Apeldoorn and A.~Gily{\'e}n, ``Improvements in quantum sdp-solving with applications,'' {\em arXiv preprint arXiv:1804.05058}, 2018.

\bibitem{durr1996quantum}
C.~D{\"u}rr and P.~H{\o}yer, ``A quantum algorithm for finding the minimum,'' {\em arXiv preprint quant-ph/9607014}, 1996.

\bibitem{dorn2007quantum}
S.~D{\"o}rn, {\em Quantum algorithms for some graph problems}.
\newblock PhD thesis, Ulm University, 2007.

\bibitem{orda1990shortest}
A.~Orda and R.~Rom, ``Shortest-path and minimum-delay algorithms in networks with time-dependent edge-length,'' {\em Journal of the ACM (JACM)}, vol.~37, no.~3, pp.~607--625, 1990.

\bibitem{lorenz2024quest}
J.~M. Lorenz, ``The quest for a practical quantum advantage or the importance of applications for quantum computing,'' {\em Modern Physics Letters A}, vol.~39, no.~21n22, p.~2430006, 2024.

\bibitem{pathak2024new}
S.~Pathak, A.~Mani, M.~Sharma, and A.~Chatterjee, ``New quantum-inspired salp swarm algorithm: A comparative study on numerical computation,'' in {\em Recent Trends in Swarm Intelligence Enabled Research for Engineering Applications}, pp.~291--330, Elsevier, 2024.

\end{thebibliography}

\bibliographystyle{plainnat} 

\end{document}